\documentclass[manuscript]{aa}
\usepackage{graphicx}
\usepackage{txfonts}
\usepackage{natbib}
\usepackage{subfigure}
\begin{document}
\authorrunning {J. Qiu et al}
\titlerunning{case study of RHESSI hard X-ray spikes}
   \title{Solar flare hard X-ray spikes observed by RHESSI: a case
    study}
   \author{J. Qiu \inst{1} \and J. X. Cheng \inst{2,3} \and G. J. Hurford   \inst{4}
          \and Y. Xu\inst{5} \and  H. Wang\inst{5}}
    \institute{Department of Physics, Montana State University, Bozeman MT
59717-3840, U.S.A.    \and  School of Astronomy $\&$ Space
Science, Nanjing University, Nanjing 210093, China
\email{chengjx@shao.ac.cn} \and Shanghai Astronomical Observatory,
Chinese Academy of Sciences, Shanghai 200030, China
 \and Space Sciences Lab, University of
California, Berkeley, CA 94720-7450 \and New Jersey Institute of
Technology,  323 Martin Luther Kind Blvd., Newark, NJ 07102,
U.S.A.}
   \date{Received  December 8 2012/Accepted September 14 2012}
  \abstract
   {Fast-varying hard X-ray spikes of subsecond time scales were discovered by space telescopes in the 70s and 80s, and are also
observed by the $Ramaty$ $High$ $Energy$ $Solar$ $Spectroscopic$
$Imager$ (RHESSI). These events indicate that the flare energy
release is fragmented.}
   {In this paper, we analyze hard X-ray spikes observed by RHESSI to understand their temporal, spectral, and spatial
properties. }
 {A recently developed demodulation code was applied to hard X-ray light curves in several energy bands observed by RHESSI.
Hard X-ray spikes were selected from the demodulated flare light
curves. We measured the spike duration, the energy-dependent time
delay, and count spectral index of these spikes. We also located
the hard X-ray source emitting these spikes from RHESSI mapping
that was coordinated with imaging observations in visible and UV
wavelengths.}
{We identify quickly varying structures of $\le$ 1 s during the
rise of hard X-rays in five flares. These hard X-ray spikes can be
observed at photon energies over 100 keV. They exhibit sharp rise
and decay with a duration (FWHM) of less than 1 s.
Energy-dependent time lags are present in some spikes. It is seen
that the spikes exhibit harder spectra than underlying components,
typically by 0.5 in the spectral index when they are fitted to
power-law distributions.

RHESSI clean maps at 25--100 keV with an integration of 2 s
centered on the peak of the spikes suggest that hard X-ray spikes
are primarily emitted by double foot-point sources in magnetic
fields of opposite polarities. With the RHESSI mapping resolution
of $\sim$4\arcsec , the hard X-ray spike maps do not exhibit
detectable difference in the spatial structure from sources
emitting underlying components. Coordinated high-resolution
imaging UV and infrared observations confirm that hard X-ray
spikes are produced in magnetic structures embedded in the same
magnetic environment of the underlying components. The coordinated
high-cadence TRACE UV observations of one event possibly reveal
new structures on spatial scales $\le$ 1--2\arcsec at the time of
the spike superposed on the underlying component. They are
probably sources of hard X-ray spikes.}
   {}


   \maketitle
%

\section{Introduction}
Solar flare emission on fine time scales of fewer than a few
seconds were first reported in hard X-ray observations from the
S-100 solar hard X-ray spectrometer aboard the ESRO TD-1A
satellite over the period from 1972 March 12 to 1973 October 1
\citep{Vanbeek74, Vanbeek76, Hoyng76, dejager78}. Numerous
short-lived hard X-ray spikes were found with the rise and decay
time as low as 1.2 s, approaching the time resolution of the
instrument. These are called ``elementary bursts"
\citep{dejager78}. A few years later, fast-varying spikes with
even shorter time scales were discovered in microwave and hard
X-ray observations \citep{Kiplinger83, Kiplinger84, Kiplinger89,
Kaufmann80, Kaufmann84, Kaufmann01, Aschwanden93, Aschwanden95,
Aschwanden97, Aschwanden98}. Using observations by the hard X-ray
burst spectrometer on the {\it Solar Maximum Mission}  (SMM) at a
time resolution of 128 ms and 10 ms, \citet{Kiplinger83}
identified 53 out of nearly 3000 flares to exhibit one or more
fast spikes with durations less than 1 s. Among these spikes, the
shortest time scale, i.e., rise and decay time and spike FWHM
duration, were several tens of ms. In recent years, efforts to
search for small-scale structures have also been made in
high-cadence and high-resolution imaging observations in optical
wavelengths such as H$\alpha$ and white light \citep{Neidig93,
Wang00, Trottet00, Kurt00}. For example, \citet{Wang00} explored
high cadence imaging observations in the far blue wing of
H$\alpha$ line and found 300-700 ms bursts at flare foot-point
kernels, which are correlated with BATSE hard X-ray peaks.
\citet{Qiu06} furthermore compared the fine temporal and spatial
structures of flare emissions observed in hard X-rays and
H$\alpha$ offband. These observations confirm that flare energy
release is fragmented, and the hard X-ray spikes are likely due to
thick-target nonthermal emissions.

Discovery of flare bursts on very short time scales have led
scientists to consider that magnetic energy release primarily
occurs on small scales. The observed fast-varying spikes may
reflect single magnetic reconnection and/or particle acceleration
events \citep{Sturrock89, Parker89, Bastian97}. The bursts' time
scales may depend on the characteristic size of the elementary
flux tubes \citep[e.g.,][]{Sturrock84, LaRosa93} or the turbulent
dynamics of the reconnecting current sheets
\citep[eg.,][]{Litvinenko96}. Specifically, spikes with very short
time scales reported in hard X-ray and microwave observations will
provide diagnostics of the acceleration of nonthermal particles in
flare environment. To unravel the physical mechanisms that govern
energy release and particle acceleration in these short-scale
events requires quality observations that push to the limit of the
existing temporal, spatial, and spectral-resolving capabilities.

The Reuven Ramaty High Energy Solar Spectroscopic Imager
\citep[RHESSI][]{Lin02} launched in early 2002 has unprecedented
resolution to further the study of the temporal, spectral, and
spatial properties of hard X-ray spikes. In this paper, we present
five flares, all exhibiting $\le$1 s structures in hard X-ray
light curves during the rise of the flares. In some events,
high-cadence imaging infrared observations by the National Solar
Observatory (NSO) and UV observations from $Transition$ $Region$
$and$ $Corona$ $Explorer$ (TRACE) are compared with spatial
structures of these spikes.

\begin{table}
\centering

 \caption{Flares with hard X-ray spikes}
\label{YSOtable}
\begin{tabular}{c c c c}

\hline\hline

  date &  start/peak/end
time  & mag. & region\\
\hline
 2002-Mar-17&19:26:16/19:29:38/19:35:48 UT&M4.0&9871\\
\hline
2002-Aug-30&13:26:20/13:29:26/13:33:12 UT&X1.5&10095\\
\hline 2003-Oct-29&20:37:36/20:47:42/21:07:52 UT&X10.0&10486
 \\
\hline
2005-Jan-19&10:14:52/10:21:10/10:31:08 UT&M2.7&10720\\
\hline 2005-Jan-17&09:35:36/09:47:06/10:38:48
UT&X3.8&10720\\

 \hline\hline
\end{tabular}

\end{table}
\section{Observations and data analysis}

The RHESSI imaging spectrometer is based on a set of nine rotating
modulation collimators (RMCs).  Their angular resolutions ranges
logarithmically from 2.3 to 183 arcseconds.  Each RMC consists of
a high spectral resolution germanium detector located behind a
pair of widely separated grids. As the spacecraft rotates (15
rpm), the grids transmit a rapidly time modulated fraction of the
incident flux so that the imaging information is encoded in the
time profiles of the detected flux.

Since the telemetry includes the energy and arrival time of each
detected photon, in principle, the time resolution for light
curves is only limited  by photon statistics;  however, the
rapidly time-varying transmission of the grids imposes temporal
artifacts of the observed count rates so that the raw count rates
from any given detector represent a convolution of solar
variations and the effect of the rotating grids. The latter occurs
on three time scales: rapid modulation (which encodes imaging
information) on a grid-dependent time scale from  4 to  500 ms;
slower modulation with a period equal to the  4 s rotation period;
and a component with half of the rotation period. The last two
terms are due to the variations in transmission of individual
grids.

\begin{figure}
   \includegraphics[width=8.5cm]{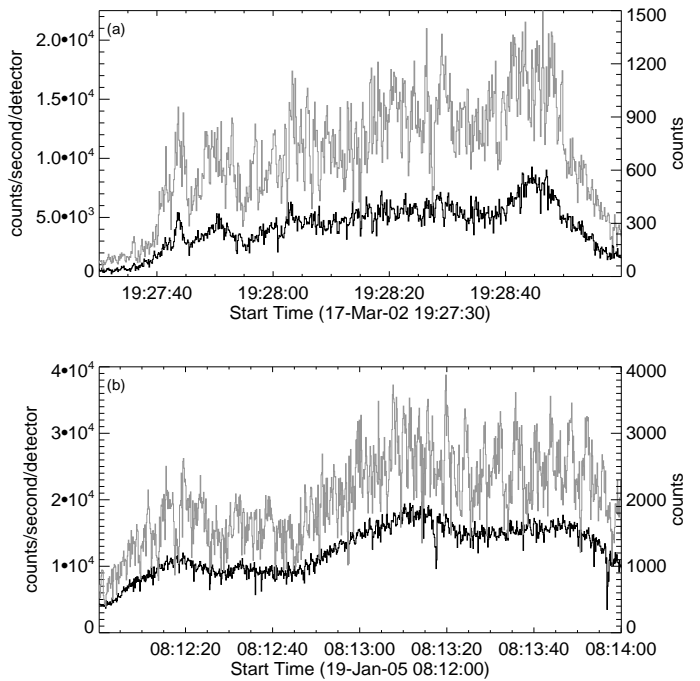}
      \caption{Demodulated RHESSI hard X-ray light curves (dark) in
comparison with summed raw data counts (gray). The light curves
are acquired at 25--100 keV. Top panel shows 2002 March 17 event
that exhibits fast-varying spikes and bottom panel shows 2005
January 19 event without significant fast-varying spikes. The
scales of the demodulated (in units of counts per second per
detector) and raw (in units of counts) light curves are indicated
on the left and right y-axes, respectively.  }
 \label{F1}
 \end{figure}

There are four known methods to suppress  the effects of
modulation (demodulation).  The first and most widely used is to
just make light curves with 4 s time resolution and so average
over all the modulation terms. A second approach is to sum the
light curves from individual detectors each of which modulates
with different frequencies or phases. This helps, but typically
leaves residual artifacts of  20\% rms.  A third alternative is an
algorithm developed by \citet{Arzner02} that can estimate a
smoothed light curve with higher time resolution, but since the
resulting data points are not statistically independent,
quantitative evaluation of temporal features in the output can be
problematic.

The demodulation algorithm used here is based on the fact that for
sources within a few arc-minutes of the (known) flare location,
the instantaneous modulation period and average grid transmission
for each RMC is both known and independent of flare morphology.
The algorithm may be used to recover time structures with the
user-chosen interval in the range of 0.05 to 0.25 s. Shorter time
limits systematically exclude additional coarse detectors, whereas
longer time limits begin to violate the important assumption that
the modulation period be constant during the interval. For each
time interval (0.125 s in this paper), the count rates in each
detector for which the modulation period is less than this
interval is fitted with the sum of a sinusoid plus a constant
offset. After dividing by the average grid transmission for that
grid and time, the offset represents a modulation-free estimate of
the (pre-grid) incident count rate over a 0.125 s interval in this
case.  Corresponding estimates from all grids with sufficiently
rapid modulation are averaged to yield the best estimate of the
incident count rates at that time. Additional minor corrections to
compensate for residual calibration uncertainties are also
applied. It is important to note that the incident flux estimates
for each time interval and for each energy interval are
statistically independent.  As we shall see, this helps with the
evaluation of temporal features in the resulting demodulated light
curves.

Figure \ref{F1} shows examples of high-cadence hard X-ray light
curves of flares that are demodulated from the raw data with this
recently developed algorithm. A large number of fluctuations are
present in the raw data set, most of which are removed in the
demodulated light curves with real hard X-ray spikes standing out.
However, hard X-ray spikes are not found  in all flares. Figure
\ref{F1}a shows an event with evident fast-varying structures,
such as the spike at 19:27:43 UT. While for the event in Figure
\ref{F1}b, no outstanding fast-varying spikes can be recognized in
the demodulated light curves. \citet{Kiplinger88} found that about
20\% SMM bursts exhibit subsecond structures.

 \begin{figure*}[!t]
\centering \subfigure[]
{\includegraphics[width=2.5in,angle=90]{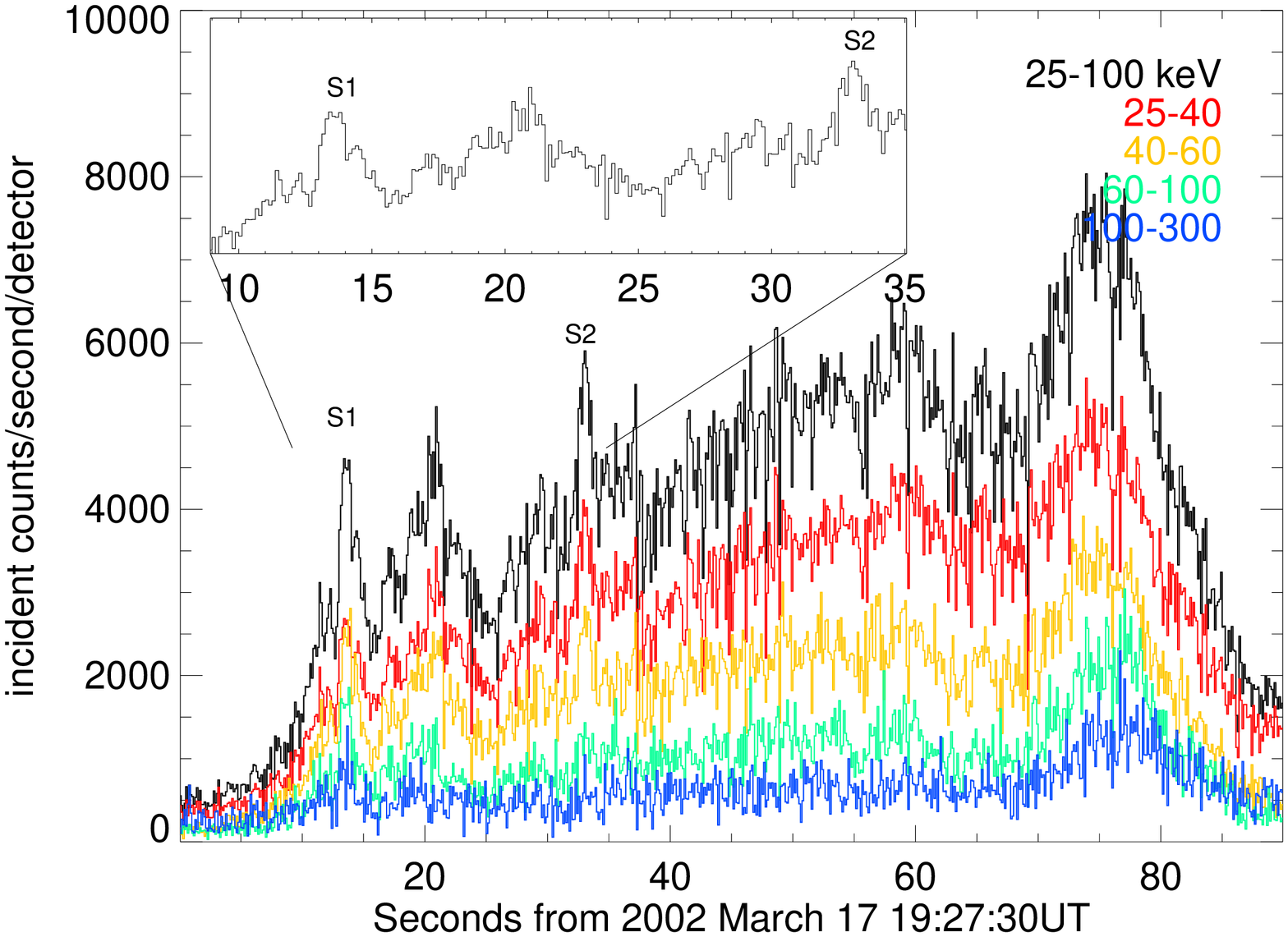}} \subfigure[]
{\includegraphics[width=2.5in,angle=90]{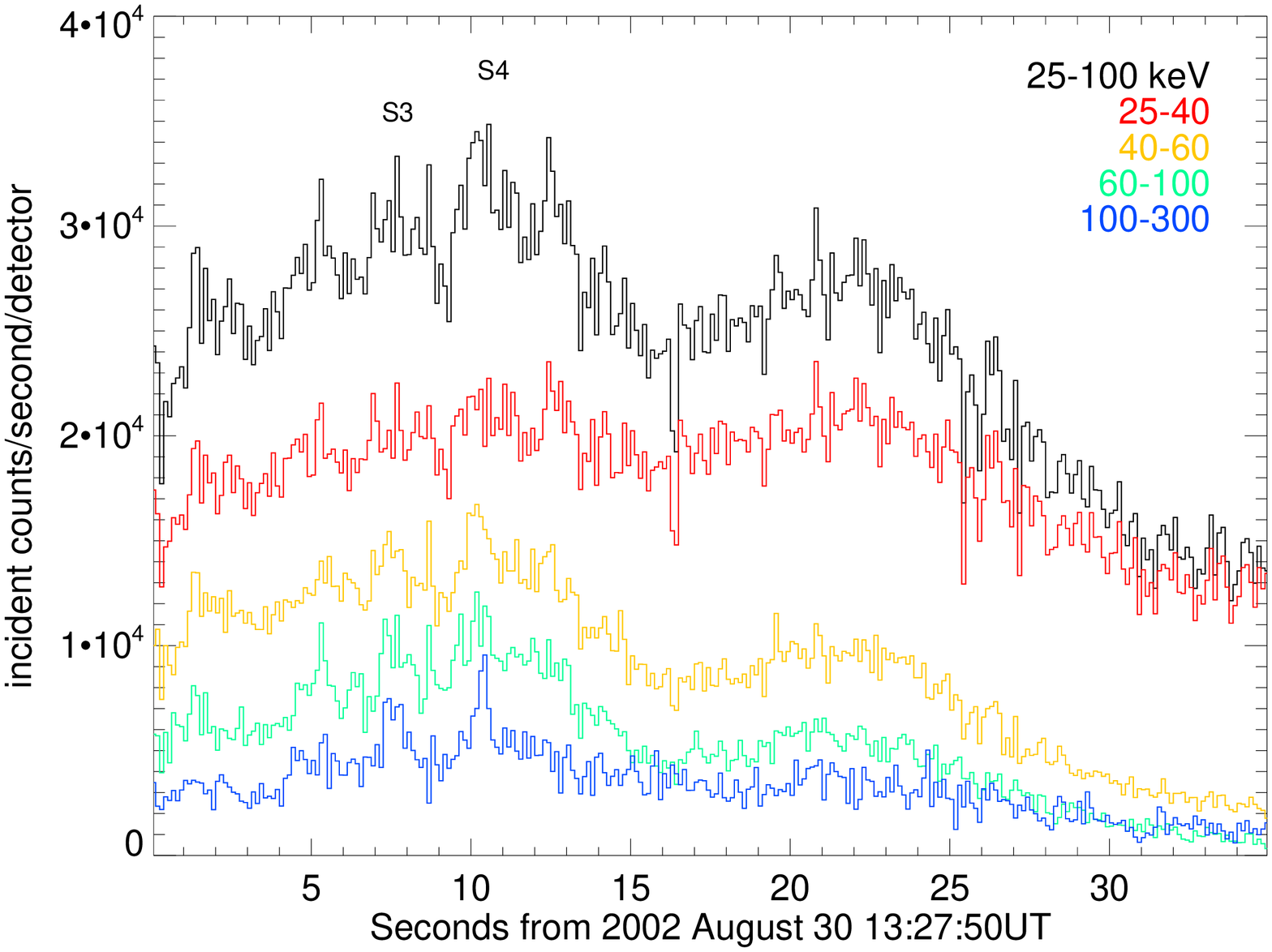}} \subfigure[]
{\includegraphics[width=2.5in,angle=90]{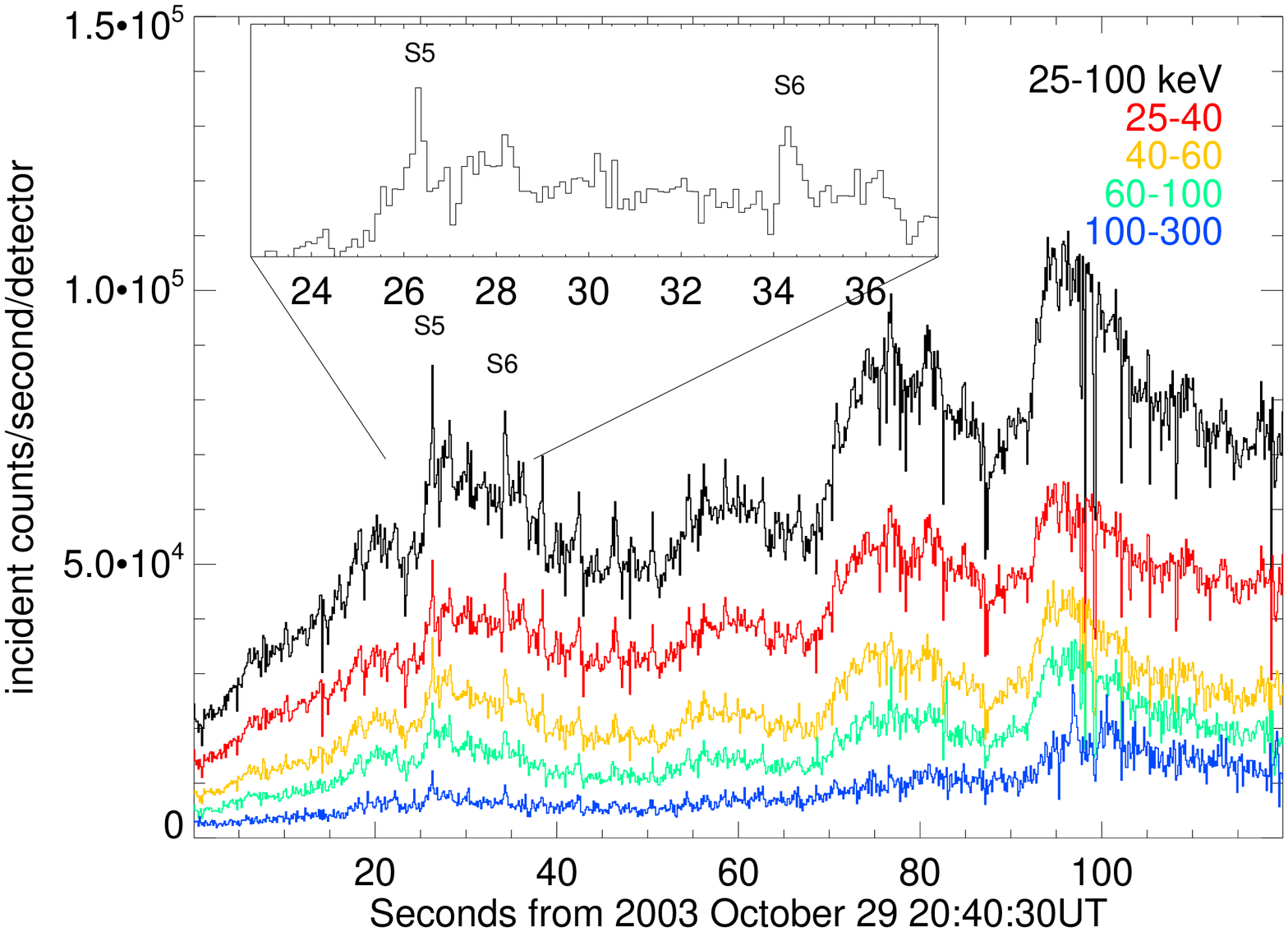}} \subfigure[]
{\includegraphics[width=2.5in,angle=90]{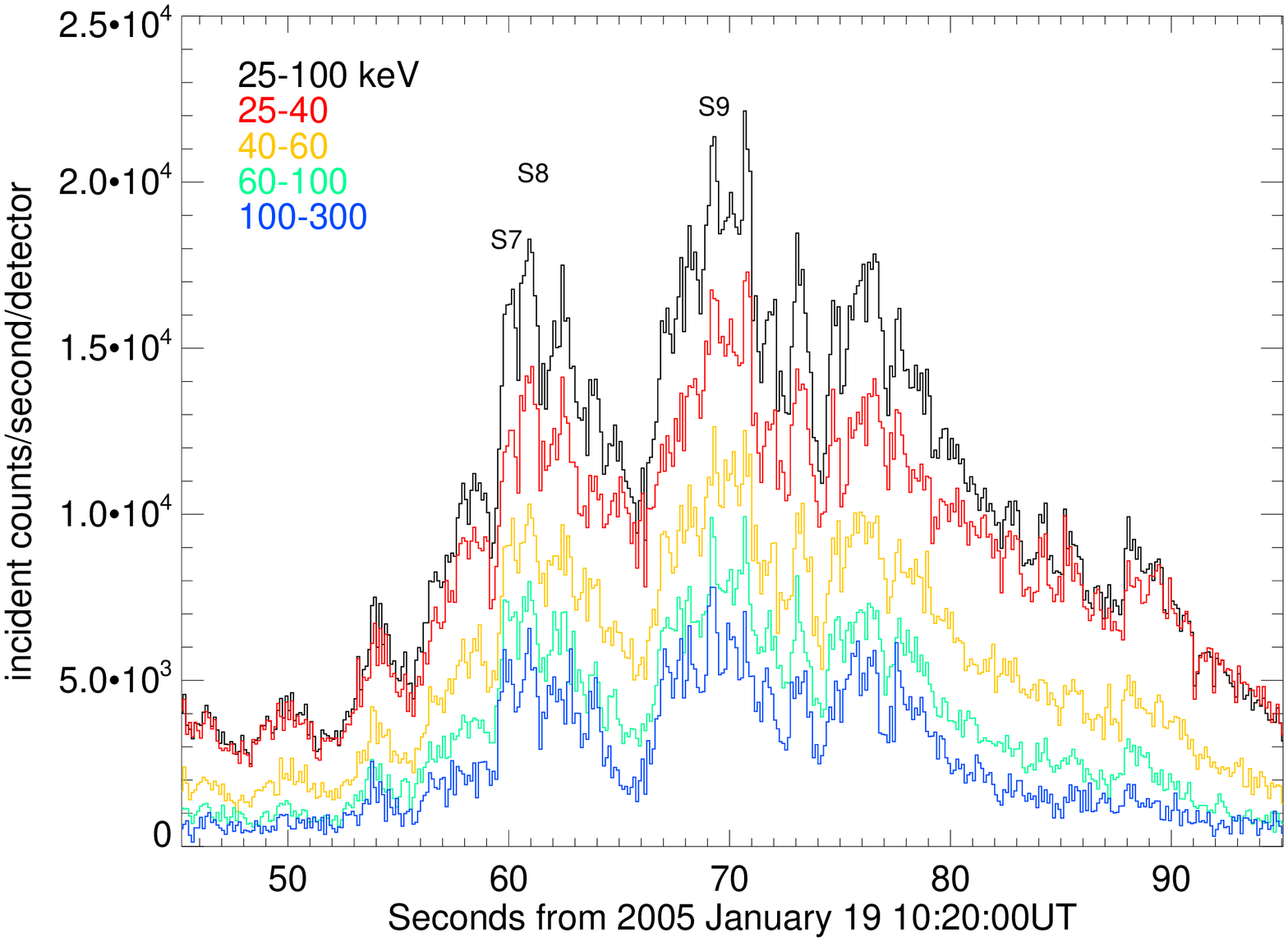}} \subfigure[]
 {\includegraphics[width=2.5in,angle=90]{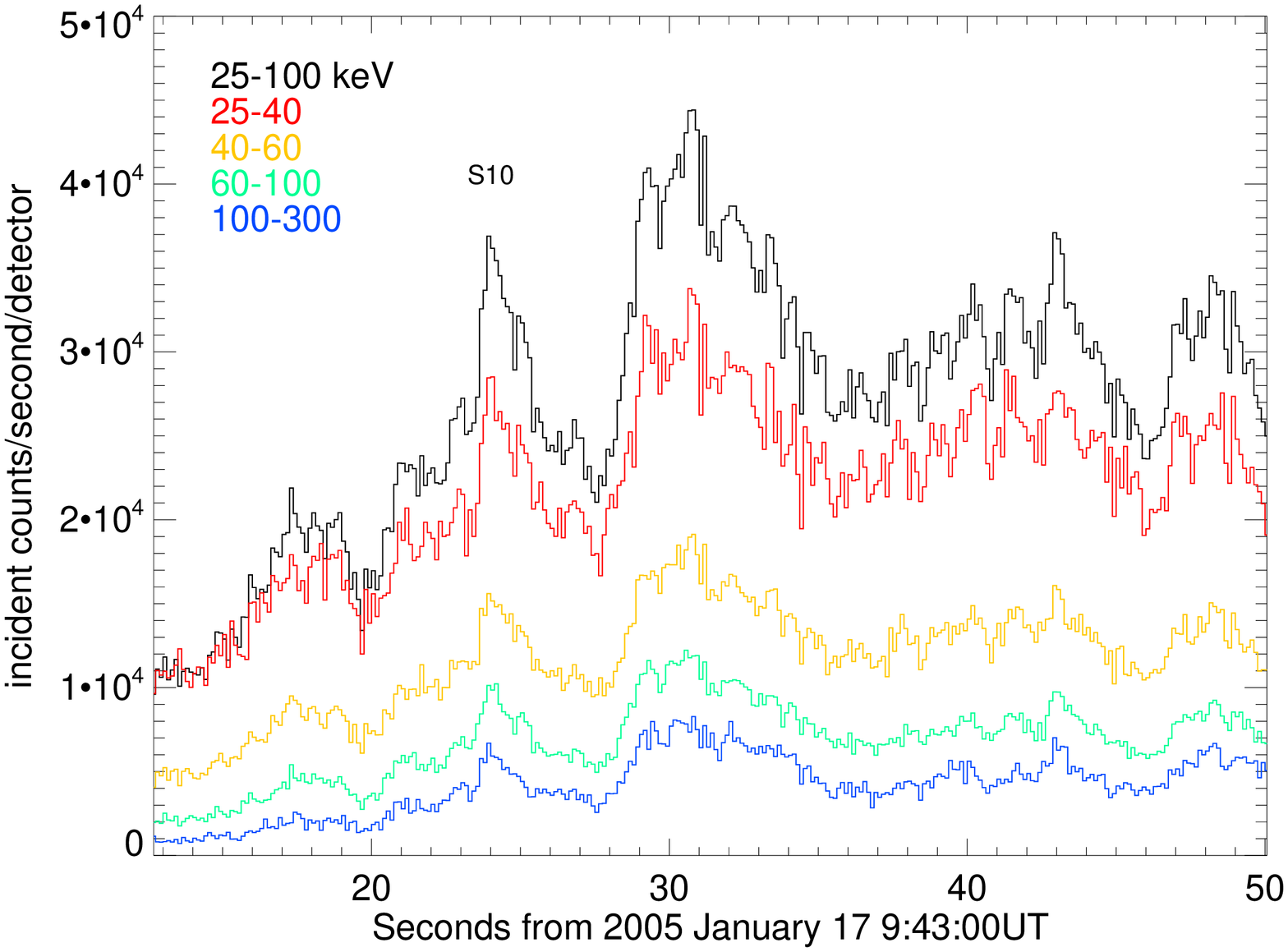}}
 \caption{Demodulated RHESSI hard X-ray light curves
at varying energies for the five flares. Then inset frame gives
magnified view of some pronounced spikes. Note that y-axis gives
the units of the 25--100 keV light curve, and light curves at
other energies are scaled arbitrarily.} \label{F2}
\end{figure*}

 \begin{table*}
\centering \caption{Properties of RHESSI hard X-ray elementary
bursts} \label{YSOtable}

\begin{tabular}{c c c c c c c c}

\hline\hline Spike & Peak time &
   \multicolumn{2}{c}{Duration (FWHM; s)} & &
   \multicolumn{2}{c}{Spectral index} &
   Source info.\\
\cline{3-4} \cline{6-7}\\

   \ \ & (at 25--100 keV) &  25--100 keV & 100--300 keV &  & spike & underlying comp. & (at 25--100
   keV)\\

\hline
S1  & 03/17/02 19:27:43.6 & 0.68$\pm$0.15 & 0.91$\pm$0.42 &  &3.07$\pm$0.40 &3.56$\pm$0.19 & double \\
S2 & 03/17/02 19:28:03.2 & 0.54$\pm$0.01 & 0.13$\pm$0.05 & &3.22$\pm$0.25 &4.88$\pm$0.48 & double\\
S3  & 08/30/02 13:27:57.7 & 0.71$\pm$0.09 & 0.41$\pm$0.11 & &2.85$\pm$0.39 & 3.48$\pm$0.35& single\\
S4  & 08/30/02 13:28:00.4 & 0.90$\pm$0.09 & 0.34$\pm$0.21 & &2.39$\pm$0.54 & 3.48$\pm$0.35& single\\
S5 & 10/29/03 20:40:56.3 & 0.46$\pm$0.18 & 0.36$\pm$0.02 & &3.10$\pm$0.14 & 3.45$\pm$0.27& double\\
S6 & 10/29/03 20:41:04.4 & 0.18$\pm$0.02 & 0.16$\pm$0.01 & &3.18$\pm$0.14 & 3.72$\pm$0.27 & double\\
S7 & 01/19/05 10:20:59.9 & 0.24$\pm$0.02 & 0.32$\pm$0.01 & &2.35$\pm$0.27 &3.29$\pm$0.23 & double\\
S8 & 01/19/05 10:21:00.8 & 0.61$\pm$0.06 & 0.29$\pm$0.01 & &2.78$\pm$0.22 & 3.29$\pm$0.23& double\\
S9 & 01/19/05 10:21:09.3 & 0.47$\pm$0.32 & 0.19$\pm$0.03 & &2.58$\pm$0.17 & 2.98$\pm$0.23& double\\
S10 & 01/17/05 09:43:24.4 & 0.93$\pm$0.45 & 0.99$\pm$0.23 &
& 2.21$\pm$0.67 & 2.27$\pm$0.67& triple\\
\hline
\end{tabular}
\end{table*}

Since analyzing RHESSI data is a time and disk-space consuming job
and it is not our primary goal in this present paper to compare
statistics with previous studies, we have not analyzed enough
events that are statistically significant. In this study, we
analyze RHESSI observations of a few flares ranging from C- to
X-class in GOES classification. Most of these events emit
impulsive hard X-rays with significant count rates ($\ge$ 100
counts s$^{-1}$ detector $^{-1}$) at 50--100 keV, making them
viable candidates for hosting hard X-ray spikes at high energies.
Analysis of these events with the demodulation algorithm shows
that five events exhibit fast-varying hard X-ray spikes, which are
visible at photon energies of $\ge$ 100 keV. The analysis results
show that hard X-ray spikes are not necessarily present in all
impulsive hard X-ray events. A more thorough statistical survey
will be presented in \citet[][; paper II]{Cheng2012}. In this
paper, we only present, as listed in Table I, the five flares
exhibiting hard X-ray spikes. We analyzed the data during the rise
phase of these events when impulsive hard X-ray structures are
most pronounced. For each of these events, RHESSI observations are
selected in a two to four minute duration without switching
attenuators and decimators for the consistency in calibration.

\section{Selection of hard X-ray spikes}
Figures \ref{F2}a--\ref{F2}e show the demodulated light curves of
the five events at varying energies from 25 to 300 keV during the
rise of the flare. The designated time bin is 0.125 s. These
events all exhibit fast-varying structures on a time scale of 1 s
or less in most of the energy bins above 20 keV. Signals up to 100
keV are significant and are likely real features. Some of these
spikes are also visible in 100--300 keV, though less significant
because of reduced counts level. The inset frames in Figure
\ref{F2} give a close look at the most pronounced spikes picked
out by a combination of visual inspection and automatic selection
algorithms.

We identify fast varying structures by subtracting a slow-varying
component ($I_s$) from the demodulated light curves ($I$), leaving
the residual signals as $I_r = I - I_s$. The slow-varying
component $I_s$ is obtained with three different methods. First,
$I_s$ can be derived as the running mean of the light curve $I$.
Second, $I_s$ is obtained by applying a low-pass filter to $I$.
Third, $I_s$ at time $t$ can be derived as an integration of
emission components before $t$, which decay exponentially,thus
$I^i_s = \Sigma_0^{i-1} I_r^{i-1}exp(-k\Delta t/\tau)$, where
$\tau$ is the burst decay time scale. The time scale of a
running-mean or low-pass filter used in this study is 4 to 8 s.
The exponential decay time scale in the third method strongly
depends on properties of bursts and the duration of analysis,
which can be adjusted to achieve a reasonable fit to the data.

\begin{figure}
   \includegraphics[width=9cm]{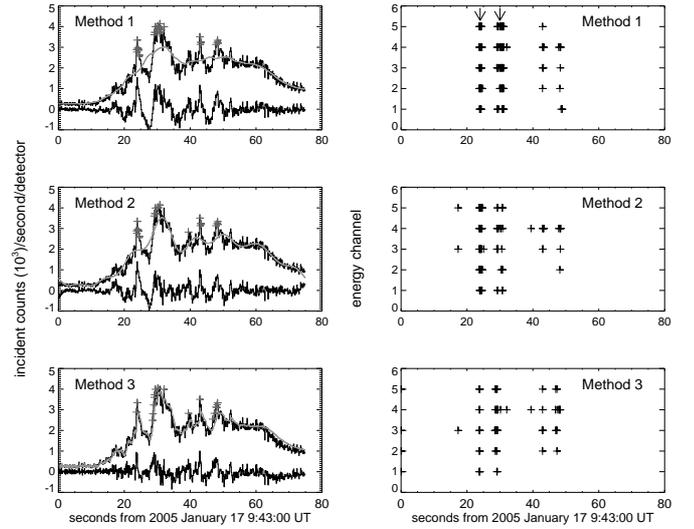}
      \caption{Example of spike detection algorithm using 3 methods
(see text) for the  2005 January 19 event. Left panels show the
demodulated light curves (dark) at 60--100 keV, the slow-varying
component (gray) determined by 3 methods, and the residuals
(bottom of each panel), with the significant residuals marked
(plus symbols). Right panels show the times (x-axis) and energies
(y-axis) of significant residuals recognized by the algorithms.
The numbers 1 to 7 along the y-axis indicate energy bands of
10--15, 15--25, 25--40, 40--60, 60--100, 100--300, 25--100 keV,
respectively. The two arrows in the top right panel indicate the
most pronounced spikes standing out in the diagrams. \label{F3} }
 \end{figure}

Derivation of $I_s$ and $I_r$ and selection of hard X-ray spikes
by these algorithms is illustrated in Figure \ref{F3} using the
  2005 January 17 event as an example. The left hand panels in the
figure show the demodulated light curve at 60--100 keV, superposed
on the slow component $I_s$ derived with the three methods. The
residuals $I_r$ are also plotted at the bottom of each panel.
Significant residuals over 3$\sigma$ level are marked in the
plots. The right hand panels illustrate the times (x-axis) and
energy bands (y-axis: see figure caption) of significant
residuals. We decide that only significant residuals, i.e., above
the 3$\sigma$ level, that span several consecutive time bins and
are present in a few, though not necessarily all, energies can be
safely regarded as real hard X-ray spikes. (However, we note an
exceptional case in the selection of S5 and S6 in the 2003 October
29  event. The demodulated light curves are full of structures,
which raises the $\sigma$ level of the residual signals,
therefore, significant residuals at S5 and S6 do not span over
more than one time bin. In this case, significant residuals are
defined as residuals over 2$\sigma$.) Abiding by this principle,
several spikes can be recognized at the times of around 24 s
(S10), 30 s, and marginally at 43 s and 48 s as well.

We applied the spike recognition algorithms to demodulated light
curves of the selected events and generated diagrams as shown in
Figure \ref{F3}. Hard X-ray spikes were picked out by visually
inspecting these diagrams. In general, we empirically required
that a spike should span at least three consecutive time bins in
at least three energy channels, and should be identified by at
least two methods. Furthermore, to derive the temporal properties
of spikes such as their rise, decay, and duration, we fit these
spikes to Gaussian profiles as shown in the following section. For
this purpose, we need to clearly separate spikes from each other
and from their background. Therefore, we discard spikes that
appear to reside in a cluster of spikes, such as the spikes at
around 30 s in Figure \ref{F3}, because they pose significant
difficulty in the fitting procedure. Eventually,  only the ten
most prominent spikes with the best Gaussian fitting results are
reported in this paper (Table II and Figure \ref{F2}).

Seen from Figure \ref{F2}, some spikes, like S1 and S2, appear to
have a rather symmetric sharp rise and decay. The time interval
between the maximum and the background in both the rise and decay
phases is about half a second, and the duration (FWHM) is below 1
s. Some of these bursts, such as S1 and S10, consist of more than
one spike each, which may be resolved if higher temporal
resolution can be reliably achieved. We also note that the most
pronounced spikes occur in the early rising phase of hard X-ray
emissions. In the maximum phase, though emission gets stronger, it
is more difficult to recognize spiky structures from the
underlying background, perhaps because of a cumulative effect by
clusters of spikes.

\section{Temporal and spectral properties of fast-varying spikes}

We investigated the temporal and spectral properties of the ten
most pronounced spikes, S1 to S10, identified in the five studied
events. The peak time, duration (FWHM), and peak count rates of
the spikes are measured in a few energy bins. We also derived and
compared the count rates of different energies, for the spikes and
their underlying components.

\begin{figure}
   \includegraphics[width=8cm]{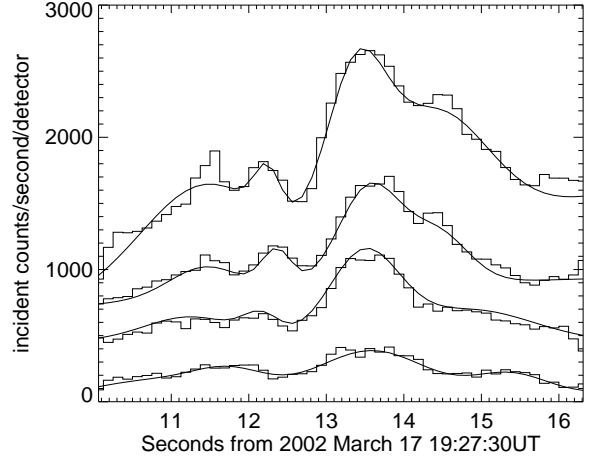}
      \caption{Demodulated light curves of spike S1 at four energy channels, 25--40, 40--60, 60--100, and 100--300
      keV from top to bottom, superimposed with the fitted curves (solid lines). Count light curves from 40--300 keV
      are arbitrarily normalized for clarity of display. \label{F4} }
\end{figure}

Since most spikes reside on top of a gradually varying background,
and many of them exhibit symmetric rise and decay, as judged from
a visual inspection of the spike light curves \citep[also see
similar conclusions by][]{Kiplinger84, Wang00}, we fit the spike
light curves in time intervals of 20 s to Gaussian profiles with a
quadratic background in the form of $I(t) = (A_0 + A_1 t + A_2
t^2) +\Sigma_i [I_i exp(-\frac{(t-t_i)^2}{2\tau_i^2})]$, where the
first three items on the righthand side represent the quadratic
background, and $I_i$, $t_i$, and $\tau_i$ represent the peak
intensity, peak time, and FWHM of each identified spike,
respectively. The Gaussian FWHM gives a measure of the duration of
the spike. Figure \ref{F4} shows an example of fitting the spike
S1 and its background at different energies.

\begin{figure}
   \includegraphics[width=9cm]{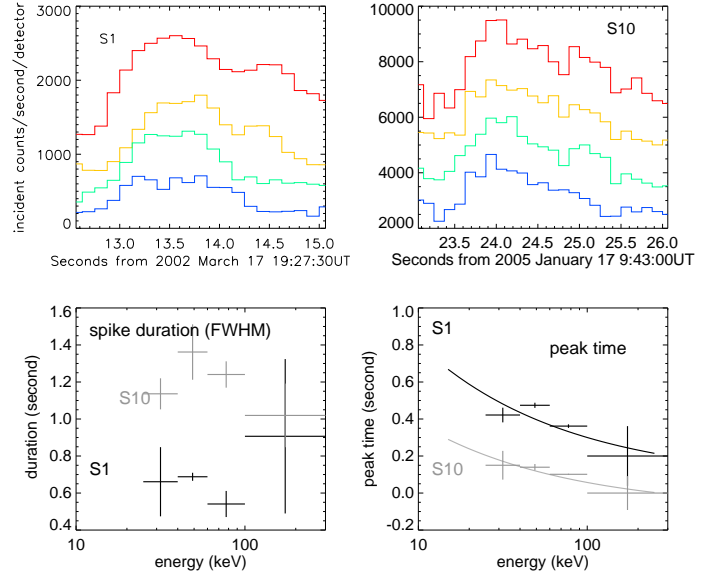}
      \caption{Examples of durations and peaking times
of the spikes S1 and S10 at varying energies, derived from fitting
the spike light curves to Gaussian profiles. Horizontal bars in
the bottom panels indicate the widths of the energy bins, and
vertical bars indicate 1$\sigma$ fitting uncertainties. Lines in
the bottom panel give the time-of-flight fit (see text).
\label{F5} }
 \end{figure}

\begin{figure}
   \includegraphics[width=8cm]{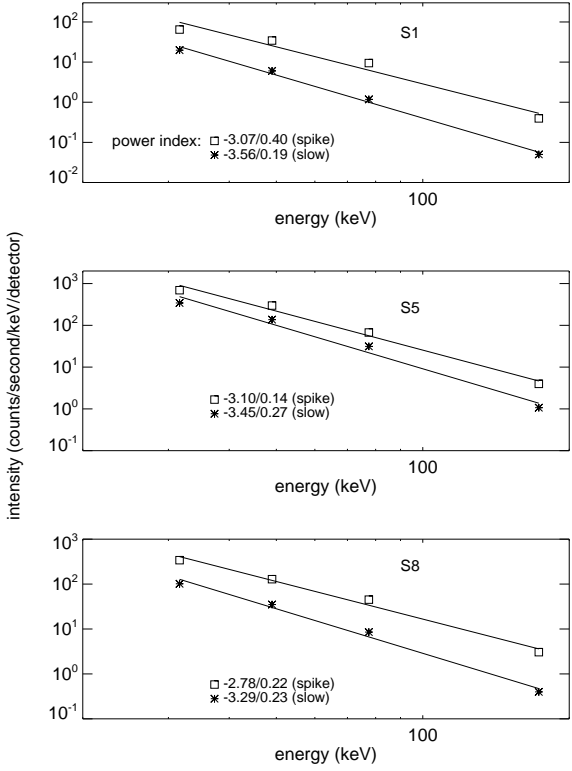}
      \caption{In each panel, the symbols and solid lines
show examples of hard X-ray photon count spectra of integrated
spike flux from the Gaussian fitting, in comparison with the count
rates spectra of underlying components. Solid lines show the
least-squares fit of the spectra to a power-law distribution, and
the exponents and their uncertainties from the fit are noted in
the figures. The y-axis gives the units of the spike intensities,
and the intensities of the underlying component are divided by
four for clarity. The three panels from top to bottom illustrate
the count spectra for spikes S1, S5, and S8, respectively, and the
underlying components. \label{F6} }
 \end{figure}

Using a Gaussian fit implies symmetric rising and decaying
profiles, which is not too far from reality, judged from a visual
inspection of the spike light curves. Earlier studies have also
revealed that the hard X-ray spikes have fairly symmetric time
profiles \citep{Kiplinger84, Wang00}.

The fitting procedure is applied to demodulated light curves at a
few energies, when applicable, in order to examine the energy
dependence of the fitting parameters. Table II lists these
parameters for light curves at 25--100 keV and 100--300 keV of the
ten spikes indicated in Figure \ref{F2}. Taking S1 and S10 as
examples, we also plot in Figure \ref{F5} the peak time and FWHM
from the fit at energies 15--25, 25--40, 40--60, 60--100, and
100--300 keV. The fitting results suggest that the detected hard
X-ray spikes have a typical time scale (FWHM) of $\le$1 s (Figure
\ref{F5}a and Table II). With the data resolution and
uncertainties in fitting, we do not find any energy-dependent
duration in hard X-ray spikes. Readers should call that the
presented time scales are an upper limit set by our selection
criterion. Spikes on shorter time scales probably exist in
reality, but they are likely to be filtered out by the selection
criterion, which requires significant residuals spanning a few
time bins. We also note that, since this paper focuses on seeking
hard X-ray spikes on very short time scales, bursts on longer time
scales are often regarded as multiple spikes residing in clusters
and are hence left out by the selection criterion.

\begin{figure}
 {\includegraphics[width=8cm]{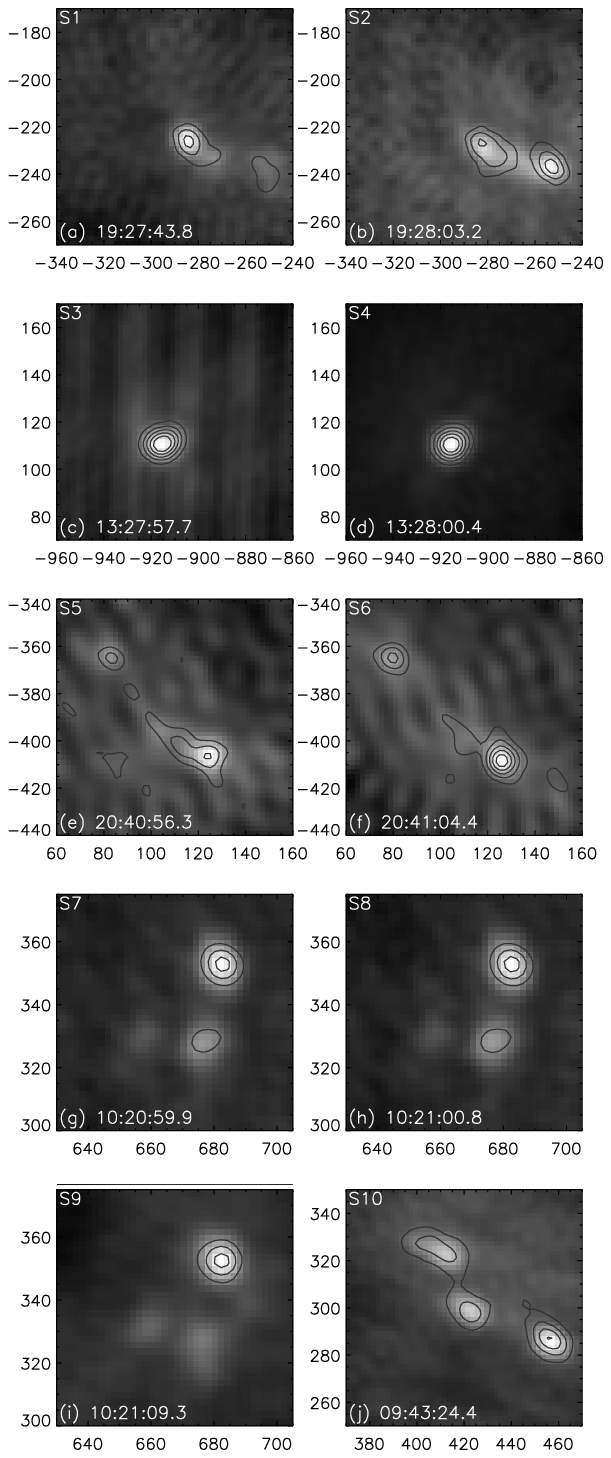}}
      \caption{RHESSI hard X-ray clean maps at 25--100
keV of spikes S1 to S10 (gray scale) superimposed on the maps of
underlying components (contours) in the same energy range. Each
map is constructed with 2 s integration. The contour levels are
0.25, 0.4, 0.55, 0.7, and 0.85 of the maximum count rates of the
image.  Axis labels indicate the E-W and N-S coordinates in
arcsec. \label{F8}}
   \end{figure}

For some spikes, e.g., S1 and S10, the fitting results suggest
energy-dependent lags in the peaking time. Figure \ref{F5}b shows
that, for S1 and S10, emission at higher energies peaks earlier.
The time lag between 20 and 100 keV photon energies is less than
half a second. If we assume that such time lags result from
time-of-flight \citep[TOF;][]{Aschwanden96} of directly
precipitating electrons, the upper limit of the loop length $L$
can be derived from the TOF measurements by $t(\epsilon) =
L(\frac{m}{2})^{\frac{1}{2}}\epsilon^{-\frac{1}{2}}$, where
$t(\epsilon)$ is the energy-dependent TOF, $L$  the loop length,
and $m$  the electron mass. For S1 and S10, $L$ is estimated to be
about 60\arcsec and 40\arcsec, respectively. These are plausible
values of coronal loop size, as RHESSI maps show the separation of
the foot points to be about 50\arcsec and 40\arcsec, respectively,
for the two spikes (see Figure \ref{F8} and Section 4). These
values are upper limits of the loop length, because realistically,
electrons spiral down along the loops with nonzero pitch angles
\citep{Krucker99}.

We further analyzed the spectra of the spikes and their evolution,
in comparison with the spectra of slow-varying underlying
components. We assumed a power-law distribution of hard X-ray
emissions, i.e., $\frac{dI(\epsilon)}{d\epsilon} \sim
\epsilon^{-\alpha}$, and make least-squares fit to the count rates
at $\ge$30 keV, as indicated by the straight lines in Figure
\ref{F6}. The fit yields the power-law index $\alpha$ of the count
spectrum. The illustrated spectra in Figure \ref{F6} are not fully
calibrated photon spectra but are count rates spectra, which,
however, can still reflect some properties, such as the hardness,
of the hard X-ray photon spectrum.

First, we fit the count spectra at the rise, peak, and decay of
each spike (not shown in the paper) to investigate evolution of
the spike spectra. Within fitting accuracies that are mostly
limited by the designated energy resolution, no significant
variation is found in the values of the power-law exponent
$\alpha$ at the rise, peak, and decay of a spike. This pattern is
different from the soft-hard-soft spectral evolution usually seen
in large bursts.

\begin{figure}
   \includegraphics[width=8cm]{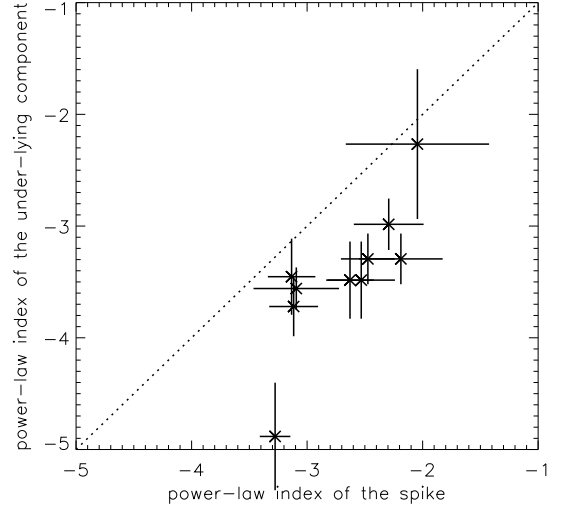}
      \caption{Scatter plot of power index ($\alpha_y$) of the spike
spectra and index ($\alpha_x$) of underlying components. The
dotted guide line outlines $\alpha_y = \alpha_x$. \label{F7}}
   \end{figure}

We then compared the spike spectra with their underlying
components. Figure \ref{F6} displays some examples of integrated
spike flux from the Gaussian fit with respect to photon energy,
i.e., the spike counts spectrum. Also plotted is the directly
measured count rates spectra of the underlying components adjacent
to the spikes. Comparison of the power-law exponents between the
spikes and their underlying components is illustrated in the
scatter plot in Figure \ref{F7}. It is seen that the $\alpha$
value derived from the count rates spectrum of the underlying
component is systematically greater than the $\alpha$ value
derived from the spike spectrum by about 0.5. This result
indicates that hard X-ray count spectrum of the spike is in
general harder than the spectrum of underlying component.

We are aware of the fact that background subtraction is key to
spectral fitting. The hard X-ray spike intensity is derived as the
integrated flux of the Gaussian component from the fit with
contribution from the underlying component removed. For the
underlying component, we take the pre-flare emission as the
background for most of the events, and post-flare emission for the
2002 August 30 event as pre-flare data are not satisfactory. The
RHESSI quick-look light curves show that during the events in the
study, the background is not subject to modulations by electron
events, thus the treatment of subtracting a mean pre-flare
background is appropriate in general. For two events, at energies
below 25 keV, pre-flare emission during the selected background
period is already high. This indicates that at low energies, the
background is over subtracted from the underlying components, and
the spectra of the properly background-subtracted underlying
components would be still steeper than illustrated in Table II.
Overall, the fast-varying spikes have a significantly harder
spectra than underlying components. This result agrees with the
general notation that hard X-ray photon or electron spectrum is
harder at emission peaks than at valleys or underlying components
\citep[e.g.][]{Kiplinger83}.

Readers are also reminded of the effect of the spike selection
principle on the distinct spectral properties between the spikes
and underlying components. As stated in the previous section, the
selection rule requires that significant residuals be present in a
few energy bands, which usually leads to preference to selecting
signals with more pronounced emission, with respect to
backgrounds, at high energies (e.g., $\ge $100 keV in this study)
than underlying components. To examine whether the data counts at
higher energies, which also carry the larger statistical
uncertainties, are crucial to the different spectral properties,
we fit the counts spectra again by eliminating the 100--300 keV
energy bin. This is a very crude test, though, since fitting
accuracies are severely degenerated in such experiment, which uses
only three points in a nonlinear fitting procedure to determine
three variables. Still, except for the  2003 October 29 event when
the exponents of the spike spectra become nearly identical to
those of the underlying components, the experiment produces
comparable or even greater difference in the power-law exponents
between the spikes and underlying components.

In general, with limitations in the spectral analysis procedures,
a conservative conclusion may still be reached that the hard X-ray
spike spectrum is harder than the spectrum of the underlying
component. The distinctive spectral properties of spikes from
underlying components may suggest that underlying components are
not an unresolved collection of fast-varying spikes. In other
words, the different spectral characters between fast-varying hard
X-ray spikes and underlying components are very likely the result
of specific physical mechanisms to generate nonthermal emissions
in such a way that the spectral properties are closely related to
the burst time scales.

\section{Imaging of hard X-ray spikes}
The spatial structure of hard X-ray spikes can be studied through
high-resolution imaging observations. In this section, we explore
RHESSI hard X-ray, TRACE UV, and NSO infrared imaging observations
to understand the morphological properties of hard X-ray spikes.

We constructed RHESSI maps of the ten spikes studied in this
paper. The adjacent underlying components are also mapped to
compare with the spatial structures of the spikes. Figure \ref{F8}
shows the RHESSI hard X-ray clean maps at 25--100 keV at the times
of the spikes. Each map is constructed with a 2 s integration
centered at the peak of the spike. Also shown are maps of
immediate underlying components. They are also constructed by
integrating data counts over 2 s immediately before or after the
spike. In principle, for maps above 25 keV, all collimators can be
used. We attempted maps including collimator 1, which is supposed
to yield the best possible resolution of 2\arcsec. However, with
the finest grid, for all events studied, stronger sources are
usually over-resolved while weaker sources are not uncovered.
Therefore, in this study, we use the collimators 2 to 9 for
optimal mapping resolution (3.92\arcsec).

In Table II, we point out the morphology of the ten spikes based
on the mapping results. Most spikes and their immediate underlying
components are primarily emitted by double sources (see Figure
\ref{F8}), likely conjugate footpoints of flare loops. The FWHM of
these sources are all close to 10\arcsec, which may be the result
of the limited RHESSI imaging resolution. The  2002 August 30
flare exhibits a single source in hard X-rays throughout the
evolution. For this event, TRACE EUV images show small flaring
loop arcade with a separation of about 10\arcsec. Given the RHESSI
image resolution, it is very likely that in the 2002 August 30
event, the spikes are also emitted by double foot-point sources
not resolved in RHESSI maps. Maps of some spikes in other events
suggest more than two sources, but the third source is usually
very weak, with the intensity about one tenth of the maximum
source intensity.

\begin{figure}
   {\includegraphics[width=8cm]{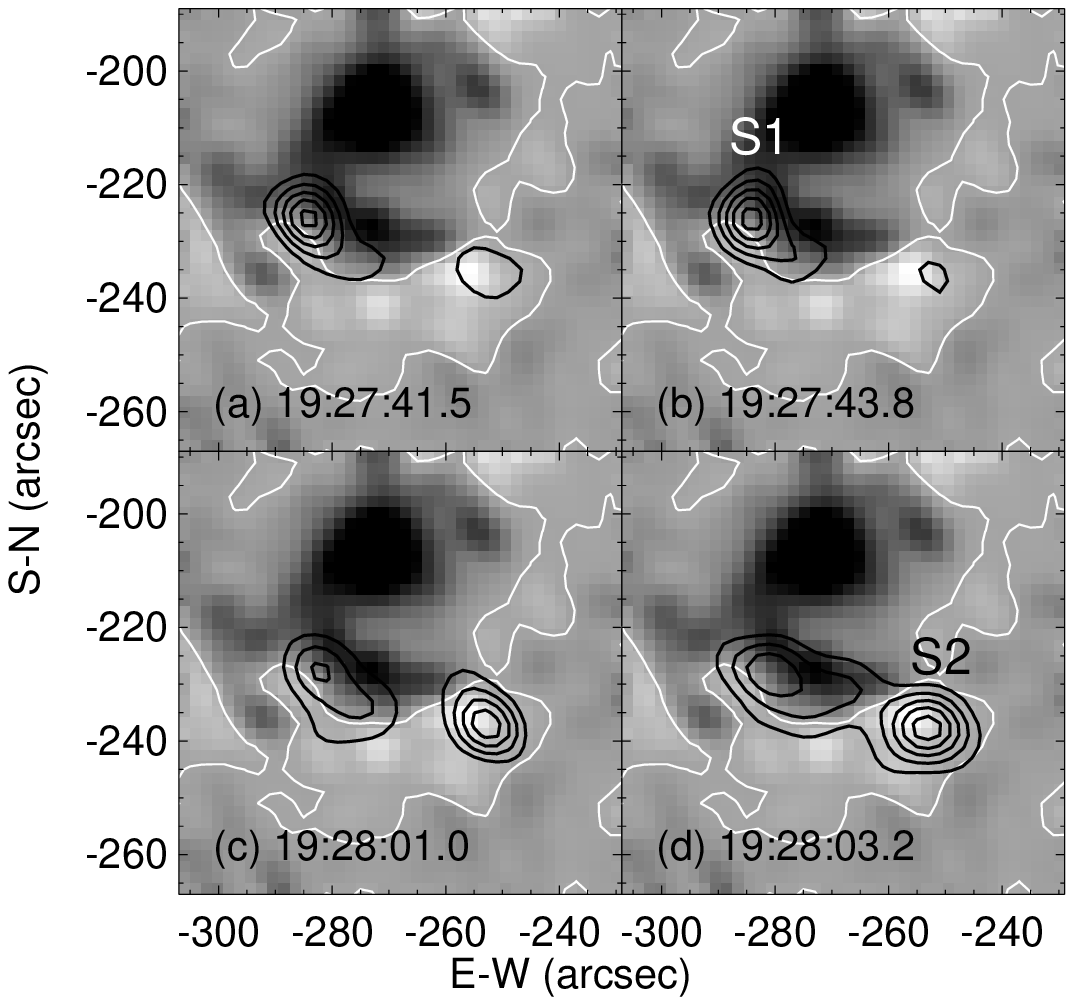}}
      \caption{RHESSI hard X-ray clean maps (dark contours) at 25--100
keV of spikes (b) S1 and (d) S2 and the pre-spike underlying
components (a and c) on 2002 March 17 superposed on the
longitudinal magnetogram taken at 19:46:01.11 UT by MDI. Each map
is constructed with 2~s integration. The contour levels are 0.25,
0.4, 0.55, 0.7, and 0.85 of the maximum count rates of all the
images, which is 0.57 photons cm$^{-2}$ s$^{-1}$ asec$^{-2}$.
White contours indicate the polarity inversion line of
longitudinal magnetic fields. \label{F9}}
   \end{figure}

\begin{figure}
   {\includegraphics[width=8cm]{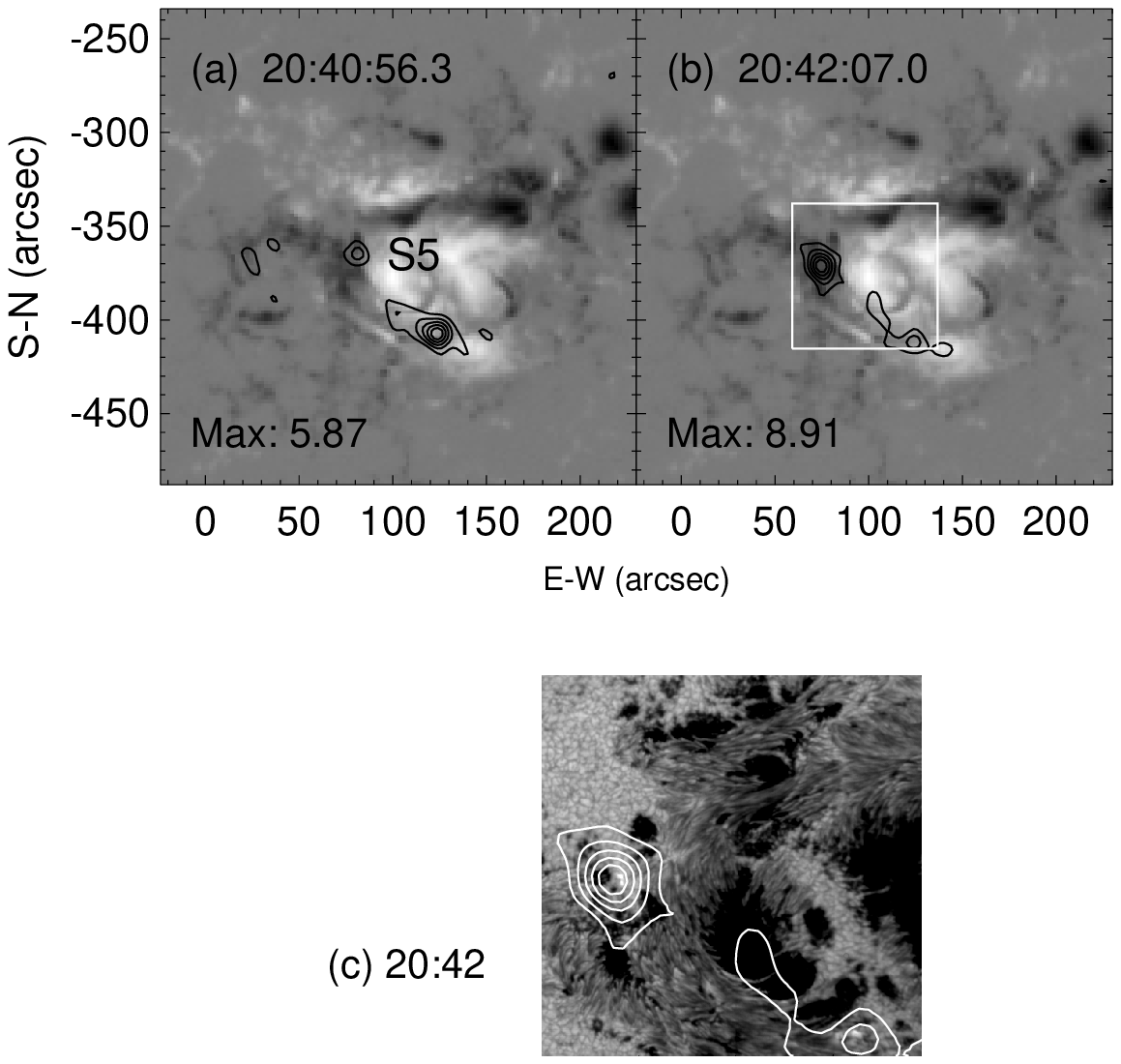}}
      \caption{RHESSI hard X-ray clean maps (dark contours) at 25--100
keV (a) of spike S5 and (b) at the maximum of the 2003 October 29
flare superposed on the longitudinal magnetogram taken by MDI.
Each map is constructed with 2 s integration. The contour levels
are 0.25, 0.4, 0.55, 0.7, and 0.85 of the maximum count rates of
each image, which is noted in each plot in the unit of photons
cm$^{-2}$ s$^{-1}$ asec$^{-2}$. (c) A snapshot of an infrared
image around the flare maximum revealing two bright kernels
coincident with RHESSI hard X-ray sources (white contours). The
white box in (b) defines the FOV of (c). \label{F10}}
   \end{figure}

\begin{figure}
   {\includegraphics[width=8cm]{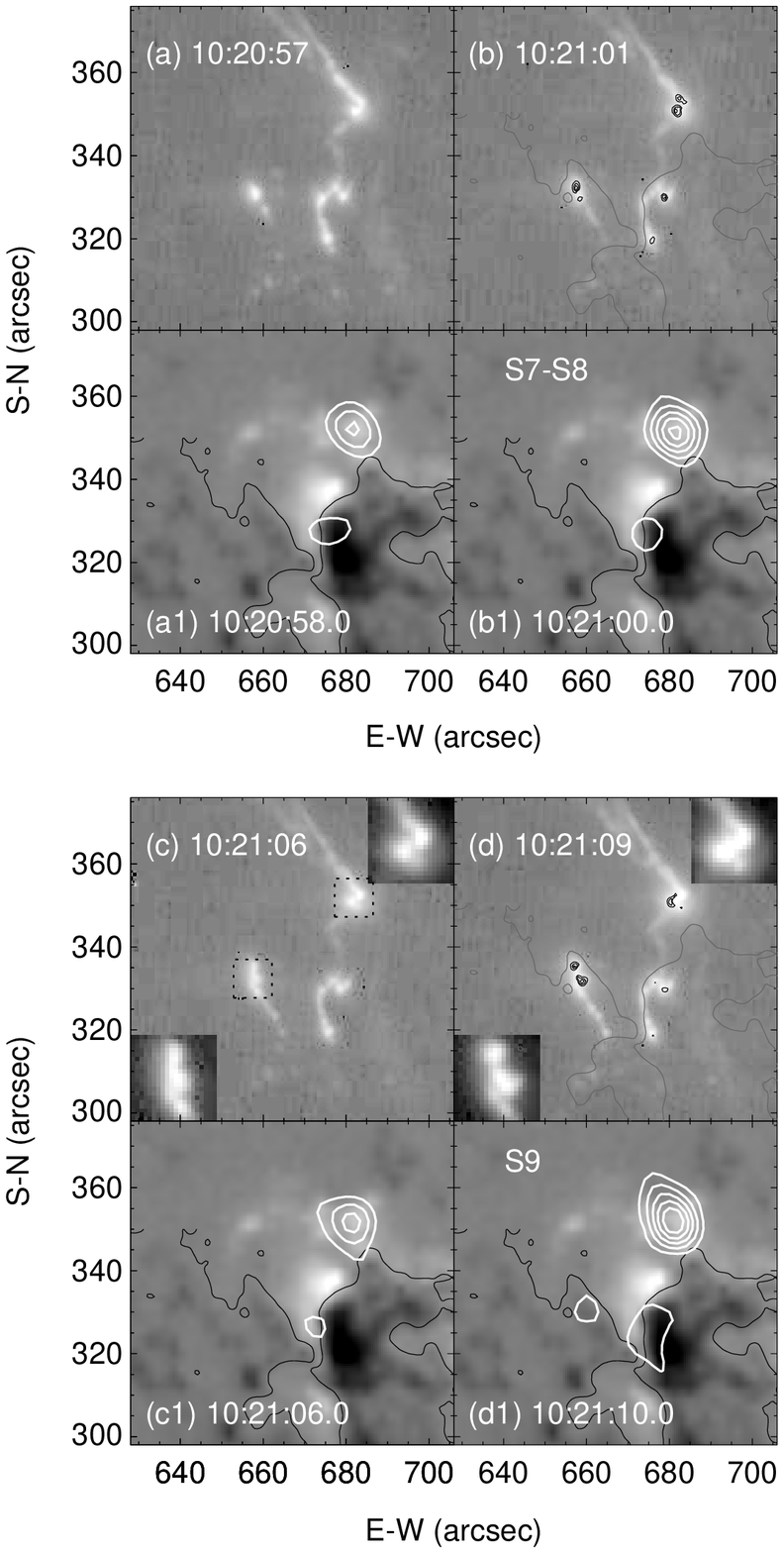}}
      \caption{Bottom panel: RHESSI hard X-ray maps (white contours) at
25--100 keV around the times of spikes (b) S7-S8 and (d) S9, and
pre-spike times (a and c) on 2005 January 19, superposed on the
longitudinal magnetogram taken at 09:36:02.32 UT by MDI. The
integration time of each image is 2 s. The contour levels are
0.25, 0.4, 0.55, 0.7, and 0.85 of the maximum count rates of the
images, which is 2.52 photons cm$^{-2}$ s$^{-1}$ asec$^{-2}$. Top
panel: snapshots of TRACE 1600 \AA\ images around the times of
RHESSI maps. The inset gray-scale images in (c) and (d) are
magnified images of two kernels, the FOV being indicated by the
white boxes in (c). The over plotted dark contours in (b) and (d)
show the difference between the spike image and pre-spike
background image by TRACE, the contour level being 0.3, 0.5, 0.7
of the maximum of each difference image. Gray contours indicate
the polarity inversion line of longitudinal magnetic fields.
\label{F11}}
   \end{figure}

To study the detailed morphology of these spikes, in Figures
\ref{F9}--\ref{F11}, we compared the images of a few spikes with
coordinated observations of the magnetic field obtained by
$Michelson$ $Doppler$ $Imager$ \citep[MDI;][]{Scherrer95}, UV
images by TRACE, and infrared images from NSO. The RHESSI images
are conveniently co-aligned with the MDI longitudinal magnetogram
using the satellite-pointing information. We then co-aligned UV
and optical images with the MDI magnetogram by registering
features, such as sunspots, pores, and plages, which are observed
in all these images. The co-alignment accuracy is limited by the
instrument resolution, which is about 1--2\arcsec.

Figure \ref{F9} shows the maps of spikes S1 and S2, and the
pre-spike underlying components for the  2002 March 17 event. S1
and S2 are both emitted by double sources about 50\arcsec apart in
opposite polarity magnetic fields. S1 primarily stems from the
source in the east, while S2 is  contributed more by the source in
the west. This is different from previous studies
\citep[e.g.][]{Wang95}, suggesting that flare footpoints in weaker
magnetic fields emit stronger hard X-rays as a result of magnetic
mirroring effects. The spatial variation between S1 and S2
suggests that the magnetic structures that emit S1 and S2 rapidly
evolve during 20 s, although they are embedded in the same
magnetic environment.

As shown in Figure \ref{F9}, the images of the spikes have nearly
identical morphologies to the images of the underlying components.
In general, for each of the events analyzed in this paper, when we
compare the RHESSI map constructed at the spike with the map of
the underlying component, no detectable difference is found in the
spatial structure between the two, although emission of the spike
is enhanced with respect to the underlying component. We note that
the RHESSI map constructed over 2 s is a combination of both the
spike source and the underlying component. As the spike duration
is less than 1 s, and the net spike emission is comparable to the
underlying component, the spike source would contribute about 20\%
to the 2 s map. This may be an important reason that the spike
source is not distinguished from the underlying component. We also
note that the RHESSI hard X-ray source size (FWHM) is around
10\arcsec, rendering it difficult to uncover the real spatial
structure of hard X-ray spikes. Alternatively, we have sought
other  high-resolution imaging observations to compare with RHESSI
data.

For the  2003 October 29 flare, high-resolution infrared
observations were taken at National Solar Observatory (NSO) around
the maximum phase of the flare. The image scale of the infrared
observations is around 0.1\arcsec, yielding a resolution
approaching the diffraction limit of 0.2\arcsec. In Figure
\ref{F10}, we compare infrared images with RHESSI maps. The
infrared observations cover only the maximum of the flare, while
RHESSI hard X-ray spikes S5 and S6 occurred a minute earlier. The
infrared images show double sources coincident with the hard X-ray
sources \citep[Figure \ref{F10}c; also see][]{Xu04} around the
maximum of the flare (Figs, \ref{F10}b and \ref{F10}c). In
infrared images, the source in the north of the sunspot can be
recognized as two patches of less than 2\arcsec in size, which
cannot be resolved in RHESSI maps. Figure  \ref{F10}a shows that
hard X-ray spikes S5 an S6 have a similar structure to sources at
the maximum, though emission in the southern footpoint is
stronger. Unfortunately, no infrared observations were taken at
the times of S5 and S6 to uncover the spatial structure of these
spikes. However, we point out that the scale of infrared emission
sources is similar to the size of a granule clearly shown in the
high-resolution images, which may be comparable to the size of an
elementary burst, as elementary bursts are thought to reflect
reconnection events between single flux tubes \citep{Sturrock89}.

For the 2005 January 19 event, high cadence (2 s) observations at
UV 1600 \AA\ were obtained by TRACE with an image scale of
0.5\arcsec, which covers the entire period of interest. Figure
\ref{F11} shows a series of TRACE UV images in comparison with
RHESSI maps at 25--100 keV. Specifically we compare the RHESSI and
TRACE images at spikes S7-S8, S9 and the underlying components.
The RHESSI maps in the lower panel in Figure \ref{F11} reveals
little spatial variation between the spike images and the
underlying component images, except for stronger emission at the
times of the spikes. TRACE UV 1600 \AA\ images in the upper panel
in Figure \ref{F11} were obtained within 1 s of RHESSI hard X-ray
spikes and their underlying components. They also show nearly
identical spike and pre-spike images, all exhibiting three bright
patches in magnetic fields of opposite polarities. Spatially
resolved UV light curves reveal enhanced emission correlated with
hard X-ray spikes at a few sites in magnetic fields of opposite
polarities, suggesting that hard X-ray spikes be emitted by
footpoint pairs rather than a single polarity footpoint or a loop
top source.

To detect the locality of hard X-ray spikes, we plot the contours
of the TRACE difference images between the spike time and
pre-spike time in Figure  \ref{F11}. For S7--S8, excess emission
at the time of the spike is exactly from the location of the peak
emission of the kernels, confirming that these spikes are emitted
by magnetic structures embedded in the same magnetic environment
of underlying emissions. For S9, the situation is slightly
different: some part of excess emission is located in new
spot-like structures extended from the edges of the background
kernels. This is most evident in the magnified images of two
kernels shown in Figure  \ref{F11}. The centers of these new
structures are displaced from the peak of the background emission
by 1\arcsec--2\arcsec, and the FWHM size of the new structures is
about 1\arcsec--2\arcsec, which should set the upper limit for the
spatial scale of the spikes if we believe that these structures
truly reflect the sources of the spikes.

To summarize, RHESSI mapping results suggest that the sources
emitting hard X-ray spikes are the same as or very closely
attached to sources that emit their underlying components, and
they are footpoint sources located in magnetic fields of opposite
polarities. This is confirmed by coordinated observations with
comparable time cadence and better spatial resolutions by other
imaging instruments. Such a result indicates that reconnection and
acceleration events that produce hard X-ray spikes take place in
the same magnetic environment of the underlying sources. Here the
word ``same" is accurate within the RHESSI mapping resolution, or
$\sim$4\arcsec.

\section{Conclusions}
We present the results in search of fast-varying hard X-ray
emissions from RHESSI observations. Spikes on time scales $\le$1 s
are found in hard X-ray emissions at photon energies from 20 to
over 100 keV during the rise of hard X-rays in five flare events.
These spikes exhibit fairly symmetric sharp rise and decay, and
the duration (FWHM) of the spike is from 0.5--1 s.
Energy-dependent time lags are present in a few spikes, suggesting
time-of-flight effects. The photon count spectra of hard X-ray
spikes are generally harder than the spectra of underlying
components, typically by 0.5 in the spectral index. This result
implies that underlying components might not be an unresolved
collection of fast-varying spikes and that hard X-ray emissions in
these events might be produced by physical mechanisms linking time
scales to spectral hardness. Future observations holding a greater
promise for precise spectral analysis should be conducted to
examine this point with more confidence.

RHESSI clean maps at 25--100 keV with an integration of 2 s
centered on the peak of spikes reveal that hard X-ray spikes are
primarily emitted by double foot-point sources in magnetic fields
of opposite polarities, which are likely superposed on, rather
than at a very different location from, the sources emitting
underlying components. With the RHESSI mapping integration time of
2 s, mapping resolution of 4\arcsec, and source FWHM size of
nearly 10\arcsec, it is hard to distinguish the spatial structure
of $\le$1 s spikes from the underlying components. Complementary
high-resolution imaging UV and infrared observations confirm that
hard X-ray spikes are produced in magnetic structures embedded in
the same magnetic environment as the underlying components, and
the scale of a spike source is likely below 1\arcsec--2\arcsec.

In spite of the limitations of existing capabilities and analysis
approaches as discussed above, our present study poses an
intriguing question: what physical mechanisms are responsible for
producing fast-varying hard X-ray spikes superposed on the
temporally and spectrally different underlying components in the
same magnetic environment? An analog to this result is the
well-known distinctive tempo-spectral behaviors between impulsive
and gradual hard X-ray and microwave bursts on much longer time
scales and usually in different flares. We also stress that the
very pronounced hard X-ray spikes reported in this paper are
conceptually different from ``elementary bursts", which the
overall flare emission is believed to be comprised of, such as the
one described in the flare ``avalanche" model
\citep[e.g.][]{Lu91}. This study is not focused on decomposing the
entire flare light curve into a sum of short-scale bursts, as
practic by \citet{Aschwanden98}. These outstanding hard X-ray
spikes are only found  in a number of flares, as first remarked by
\citet{Kiplinger83}, which indicates of specific physical
environment or mechanisms that generate them. To explore this
subject further, demodulated RHESSI observations aided by
substantially improved spectral and imaging resolving capabilities
or techniques \citep[e.g., the visibility imaging by][]{Hurford05}
are needed. The demodulation algorithm itself can be improved with
the explicit calculation of statistical errors and more
sophisticated normalization corrections. Study of the particle
acceleration associated with fast-varying spikes may be conducted
further with the next generation of hard X-ray and/or microwave
observing facilities such as the Spectrometer/Telescope for
Imaging X-rays (STIX) and Expanded Owens Valley Solar Array
(EOVSA).

\begin{acknowledgements}

We thank the referee for constructive comments that helped improve
the presentation. We acknowledge Mrs. Nora Harrington for editing
the language. This work is supported by NASA grant NNX08AE44G, NSF
grant AGS-1153424, and NSFC under grants 10878002,
 10933003, 11133004 and 11103008 and by the Chinese Academy of Sciences (KZZD-EW-01-3).
JQ acknowledges the hospitality of RHESSI group at the Space
Science Laboratory in University California at Berkeley, where
part of the work was conducted during her visit.

\end{acknowledgements}

\end{document}